# Digital Labour in a Turbulent Era: Introduction to the Book

**Paola Tubaro**

***Centre of Research in Economics and Statistics (CREST), CNRS-ENSAE- Institut Polytechnique de Paris***

***ORCID 0000-0002-1215-9145***

## ABSTRACT

This chapter introduces the book Digital Labour in a Turbulent Era. It seeks to illuminate the sources and effects of the turbulence that surrounds digital labour, seen as an expression of the antagonisms that characterize contemporary capitalism, dominated by information and communication technologies. Its central claim is that digital labour is the mirror and symbol of four broad societal tensions: incessant technological innovation, continuously reshaping the role of the 'humans in the loop' that sustain its creation and functioning; globalisation in its late phases, with its effects on countries' development; misrecognition of workers' abilities despite their key role in today's knowledge-intensive economy; and increasingly profound social inequalities. Beyond definitional debates, the chapter presents the contents of the book as efforts to concretely explore the diverse organizational forms of digital labour and its intertwining dimensions throughout the four ambits just identified. It concludes by drawing the main implications and policy recommendations.

## KEYWORDS

Digital labour; Digital capital; Platform work; Algorithmic management; Artificial intelligence; Public policies

## INTRODUCTION

Labour relations are facing extreme turbulence. Debates about the 'future of work' emphasize technology-enabled mobility, flexibility, and entrepreneurship, but also wave the spectre of automation-induced unemployment, job insecurity, and vanishing social protection. This multi-faceted and fast evolving context is epitomized by the rise of digital labour as emergent model of contingent work in the wake of technological disruption.

Originally rooted in Marxist political economy, the term digital *labour* appears as the counterpart of digital *capital* in a society in which the accumulation of resources in the economy, of power in politics, and of legitimacy in culture, are organized through technology (Fuchs, 2014; 2025). Digital labour is not just the human activity that produces specific types of goods, but an expression of the antagonisms that characterize the current phase of capitalism, dominated by information and communication technologies. It is indicative of today's turbulent times in which relentless technical change renews tensions between capital and labour. For example, the rise of digital platforms as job market intermediaries (Casilli & Posada, 2019), the growing demand for human inputs in the global value chains of artificial intelligence (AI) (Muldoon et al., 2026), and the diffusion of algorithmic management across workplaces (Aloisi & De Stefano, 2022) challenge established employment relations, once-shared understandings of the meaning of work, and the rights of workers. Indeed, concerns that digital technologies generalize precariousness and low pay attract growing policy attention (ILO, 2021).

It is because of the turbulence that it brings to the world of work writ large, that the concept of digital labour has gained prominence in scholarly and public debates – even though it may have seemed a niche concept when it was first created to signify the unpaid value-producing activities of social media users (Terranova, 2000; Scholz, 2012). Over time, it has come to encompass a broad range of non-standard work arrangements mediated by technologies. Today, it is commonly understood as referring to platforms like Uber and Deliveroo, which use data and algorithms to match clients with workers, construed as independent contractors, for one-off 'gigs' without any long-term commitment. But platform labour is not limited to these familiar instances in which tasks are carried out in person in given physical locations, like transport, delivery, and personal care services: there is also a less mediatized, but nonetheless huge (with several million workers worldwide according to Datta et al., 2023) online labour market in which work assignments are performed remotely, from graphic design (Longo & Fernández Massi, 2023) to data entry and even just 'liking' and 'sharing' contents on social media (Lindquist, 2022). What's more, online and offline platform intermediation does not exhaust digital labour. If lively debates accompanied the initial focus on unpaid user activity, at the other end of the spectrum the notion can be stretched to cover various forms of salaried employment, like the providers of data services in the global value chains of AI (Casilli, 2025), and the formal employees who generate value for digital media industries (Jarrett, 2022). Some authors go as far as to include the factory workers who produce hardware (Qiu, 2017), and the miners who extract minerals for it (Fuchs, 2014). Telework is often thought to be in scope too.

This evolution has fuelled turbulence in the literature too. Gandini (2021) warns that the breadth of many current uses of the concept weakens the original critical Marxist stance on labour and value. Conversely, narrower definitions may be too partial and occlude from view the new contours of the (digital) working class, or the conditions for labour struggles. Two recently published handbooks take sides with the latter, more comprehensive perspective. One defines digital labour very generally as a 'technologized condition of work, characterized by vastly different ownership schemes, labor processes, and reward systems' (Qiu et al., 2025: xix). The other seeks consensus through a pragmatic definition of digital labour as 'an interdisciplinary field of inquiry, a space where scholars of various disciplines and orientations converge to explore the intersections of labor and digital technologies' (Bulut et al., 2025: 2). This approach is by no means less ambitious, as its goal is to stimulate academic cross-fertilization and joint efforts 'to shape changes within and beyond workplaces' (Bulut et al., 2025: 2).

This book takes these insights on board and engages with the multi-faceted nature and growing scope of digital labour at the crossroads of the technological, societal, political, and economic transformations that shape our present times. It seeks to illuminate the sources and effects of the turbulence that surrounds digital labour, seen as the mirror and symbol of broader societal tensions in the current phase of capitalism. Moving beyond still fundamental, but somewhat disembodied definitional debates, it explores the concrete diversity of organizational forms of digital labour and brings attention to its intertwining dimensions. Without claiming to be exhaustive, it aims to offer a synthesis of the state of the art and to open avenues for further (theoretical and empirical) investigation, across various scientific disciplines and countries.

## FOUR LINES OF INQUIRY

A central claim of this book is that the turbulence is magnified at the intersections of digital labour with other major trends that shape today's societies: not only incessant technological innovation, but also the late phases of globalisation and their effects on world development, misrecognition of workers' skills and education affecting their careers and employability, and increasingly profound social inequalities. These four domains were less central in the early literature and have emerged recently, as non-standard work arrangements fomented by technologies have been woven throughout broader socio-economic transformations.

### The labour-technology nexus

A primary source of turbulence is technology itself, notably the fast-paced development of AI. While public debates overwhelmingly foreground the potential unemployment that may result from automation, the very production of AI also creates jobs – from the highly-paid engineers and computer scientists who design 'smart' systems to myriad lower-level support positions in areas such as content moderation (filtering out violent, illegal, or otherwise disturbing text, images or videos), data annotation (labelling unstructured text, images or videos so that machine learning models can detect patterns from them), and output verification (checking the accuracy of a model's results: for example, of a chatbot's responses to a user's prompt). These activities go under the name of 'AI data work' (also known as 'micro-work', 'crowdwork', and 'ghost work': see Muldoon et al., 2024). They showcase a paradox that has attracted significant scientific and policy attention: these new jobs come at the price of low pay, employment insecurity, and lack of social protections (see e.g., Miceli et al., 2020; Miceli & Posada, 2022).

AI is also used for algorithmic management, which can be defined as the delegation of decisions to automated systems (Aloisi, 2024). Digital labour platforms have served as testing tools and spaces for experimentation of new forms of algorithmic management, automating the matching of actors on opposite sides of the market, at scale – thereby controlling workers even without hiring them as salaried employees (Rani & Furrer, 2021). The lessons learned from these experiences are even more valuable as automated systems are being deployed in conventional workplaces too – for tasks that range from job candidate selection to employee surveillance.

However, delegation of responsibilities to automated systems is almost always partial: just as AI systems need inputs from data workers to be built, some 'humans-in-the-loop' also intervene, more or less transparently, to complete, correct or direct algorithmic operations. A range of expressions, from 'fauxtomation' (Taylor, 2018) to 'deceptive AI' (Rani & Dhir, 2024) and 'impersonation' (Tubaro et al., 2020), denote the ubiquitous practice of concealing the human contribution to inflate the merits of an allegedly automated system. In fact, the apparent conflict between humans and machines is only a cover for the resurgence of the more traditional power struggles that oppose different human groups underneath. Who chooses the inputs that feed AI? Who has control over algorithms, and who, instead, endures the consequences of their decisions? How do different groups endeavour to turn algorithmic decisions to their advantage? How do they resist (if at all)? In sum, the relationship between labour and technology goes both ways – one (re)shapes the other, in ways that require further investigation.

## Digital labour in a globalized world

While early studies scrutinized digital labour in North America and Europe, recent research uncovers its forms in the so-called Global South (or Majority World). Platform delivery workers now punctuate the cities of, among others, Argentina and Chile (Gutiérrez-Crocco & Atzeni, 2022); the value chains of AI extend to Latin America, Africa, and South-East Asia (Casilli et al., 2025); the Philippines have become a global hub for content moderation (Roberts, 2019); and African designers, programmers, and writers (Anwar & Graham 2022) offer their services to clients overseas. In countries in which job markets are stagnant, digital labour is often seen as an improvement compared to alternatives in the informal sector. Especially in crisis-stricken economies like Venezuela, platform-mediated online labour has become a highly sought-out source of hard currency – a way of providing remittances without physically migrating (Torres Cierpe & Tubaro, 2025).

One may thus be inclined to see digital labour as a welcome earning opportunity in otherwise sluggish economies. Likewise, platformisation may appear as a first step in the transition from informal to formal employment. But these short-run effects may be insufficient to sustain a longer-term process of development. Pressure to keep labour costs down is a key reason for platforms and business process outsourcing (BPO) operations to move to the Global South, but it implies low wages, short and sometimes even inexistent contracts (thereby bringing informality back in: Surie & Huws, 2023), and weak or absent social protection. Thus, digital labour may not fully cover the living costs of workers and shift part of them onto their local communities, as shown by Posada (2022) in the case of Venezuela. Adapting an idea that Schor (2020) formulated for US data, platforms and BPOs supply cheap digital labour to richer regions by free-riding on the local economies of the Global South. To what extent, then, is the current turbulence inherited from the asymmetrical relationships of extraction observed in earlier waves of globalisation? Is digital labour a signal of the worldwide diffusion of the informal conditions already prevalent in the South, rather than a newfangled phenomenon? Under these conditions, can a global digital working class be distinguished?

## Misrecognised abilities in the knowledge economy

Anyone – so goes the usual account – can post a message on social media, ride a bike to deliver a meal, and tag objects in a photograph. Put differently, digital labour is commonly construed as an unskilled activity, which therefore deserves low pay. Yet it requires at least some digital literacy – a resource that in standard job markets, typically brings advantages (Hargittai, 2021). Digital labour is fully part of the knowledge economy and contributes to it through the production of data that all workers undertake, either directly as their primary task (for example, data annotators) or indirectly as a side activity needed to support their main one (for example, drivers who curate their profile by adding information that may attract customers) (Casilli, 2025). From an empirical perspective, many workers have high levels of educational attainment (ILO, 2021) and actively engage in workplace learning, even without employer-provided training (Margaryan et al., 2022). If task execution itself does not always require special abilities, workers must become able to navigate the world of digital labour: to scan the competition, to guess the needs and preferences of clients, and to use time efficiently.

This apparent inconsistency is partly the result of the crowding-out effect that occurs when in conditions of uncertainty, skilled-yet-impoverished middle-class workers grab labour opportunities that were traditionally left to the working class, as in transport and delivery (Schor, 2017). It may also be the product of algorithmic management, which replaces educational credentials with reputation scores and performance trackers in ways that often disorient workers. Either way, the observed gap sits at the intersection of conflicts – between categories of workers, and between labour and capital – and results in widespread misrecognition of workers' abilities, goals and achievements. New research foregrounds intersubjective recognition as a foundation of practical identity and a site of struggle when individuals or groups are denied it – notably when platforms and other digital environments disrupt this process (Newlands, 2022; Tubaro, 2022; Wood & Lehdonvirta, 2023).

Better understanding the dynamics of (mis)recognition is important not only to make sense of digital labour today but also to assess its long-term effects. If the experience of (say) platform delivery could serve as a springboard to better jobs later, then the overall effects of digital labour would be good at least for some of the workers. But if this experience is not recognized as valuable, then these positive impacts may never materialize. What occupational opportunities do workers pursue after digital labour? How do they see their own skills? How do they leverage the learning undertaken in platform contexts? In short, what does a career look like for a delivery courier, a driver, a data annotator, or a content moderator? Because the phenomenon is recent, most extant studies are snapshots at specific moments in time. There is a need for more research, theoretical and empirical, to illuminate workers' strategies and trajectories over time.

## Social inequalities

Access to computing equipment, internet connection, and at least minimal numeracy is increasing worldwide, but it is still not granted to all and continues to separate the digital-haves and have-nots (Robinson et al., 2020a), especially in lower-income countries. The paradox of digital labour is that workers are on the right side of the digital divide, but on the wrong one in terms of job security and income. So, do digital resources backfire? A more plausible explanation harnesses the idea of the 'digital inequality stack', designed to capture the complexities of digital inequalities as they occur on many interrelated layers and work together (Robinson et al., 2020a). The foundational layers of the stack refer to devices and access to network infrastructure, while higher layers involve knowledge, skill building, differentiated use and production. The former, which digital labourers largely have access to, are necessary but not sufficient to enable the latter. In this perspective, new technologies – from platform-based matching to algorithmic management – can simultaneously increase access and connectivity, while also generating new avenues for the reproduction of disparities at higher levels. These recently theorized 'emerging inequalities' (Robinson et al., 2020b) require further exploration.

But within digital labour, are workers all equal? Extant research has often lumped them in a single category to stress the loss of labour rights and protections in comparison to conventional employees. This perspective masks differences among workers, and only recently has it become clear that, for example, platform labour is predominantly undertaken by immigrants and/or by racial and ethnic minorities, less likely to find alternatives in the formal economy and less protected even in the case of a hypothetical reclassification (van Doorn et al., 2023). Likewise, women are less present than men, especially in developing countries (ILO, 2021), and even in online freelancing and data work, which can be done from home, they earn less (Adams et al., 2025; James, 2022). What are the socio-economic mechanisms that attract particularly disadvantaged sub-populations toward digital labour, and why do they continue to engage with it even when its shortcomings become evident? Conversely, what are the mechanisms that limit women's involvement even when flexibility and work-from-home could make digital labour attractive for them? These questions call for social-scientific, but also for legal and institutional, responses: are existing anti-discrimination norms sufficient to make the digital labour market fairer? What other regulations can be leveraged? Is there a need for specific, new legislation?

## CONTENTS OF THE BOOK

The chapters that compose this book explore the boundaries of digital labour in each of the ambits just outlined, comparing its manifestations and implications in varying settings (food delivery, urban transport, AI data work, creative freelancing, telework) and geographies (Europe, North America, Latin America). This fourfold structure does not prevent chapters from resonating with one another across parts, defining multiple guiding threads: for example, algorithmic management and platform infrastructures are dealt with in several chapters which, despite differing approaches (conceptual/empirical, case-based/general) consistently portray them as socio-technical – rather than purely technical – objects.

### The labour-technology nexus

The three chapters that compose this section unpack the relationship between labour and technology in two ways: they reveal the human contribution underpinning seemingly automated decisions and showcase how algorithms embed power relationships – notably between capital and labour. First, Iraklis Vogiatzis and Manolis Patiniotis uncover the computing tools and technical solutions through which workers are transformed into ‘functions’ within hybrid human-machine assemblages. Focus is on how machines’ ‘micro-tasking’ enabled humans’ ‘micro-work’ (today, also referred to as ‘AI data work’ as mentioned earlier) through platforms in the early 2010s – a founding moment that durably established the linkages of digital labour with AI. This case clearly highlights how the goals of capital accumulation are embedded into the technical architecture of technology production.

Milagros Miceli and Adio Dinika continue to explore the human labour that feeds AI, this time foregrounding an epistemic perspective. This labour serves to build ‘ground truth’ for models, but workers' situated knowledge does not participate to it. Specific viewpoints are imposed to them through strict task instructions, pre-defined taxonomies, and surveillance-driven quality control. Models reflect the ‘truths’ of the powerful – and are neither neutral nor self-evident.

In the following chapter, Francisca Gutiérrez Crocco challenges deterministic accounts of algorithmic management as all-powerful and fully automated. In the case of food delivery platforms in Chile, she reveals the (manual) contributions of multiple stakeholders behind the scenes – from couriers to support staff and clients. Algorithmic power is continuously disrupted and restored through the struggles and negotiations of everyday human interactions. Thus, any regulation requires a broad framework encompassing both technical and social components.

**Digital labour in a globalized world**

The second part positions digital labour in the political-economic context of contemporary capitalism, shaped by successive waves of globalisation and evolving class struggles. It opens with Julian Posada’s conceptualisation of digital platforms as infrastructures of extractivism, through which companies from core AI production hubs (like Silicon Valley) appropriate data and labour from peripheral areas, notably in the Global South, in an unequal exchange. By facilitating the outsourcing and offshoring of data tasks at planetary scale, platform technologies maintain a transnational system of accumulation, labour exploitation, and power concentration that originated in colonial time and perpetuates its legacies.

Can developing countries move from extractive relationships to a steady and sustainable growth path? Uma Rani argues that vigorous state-led policies would be needed. Digital platforms are insufficient, as they merely absorb surplus labour into precarious roles that fail to create the infrastructures and the capabilities essential for a genuine economic take-off. The transition requires ambitious public investments and the strengthening of social and economic institutions to ensure that gains are retained locally and distributed equitably across the economy.

Then, Matheus Viana Braz zooms in on digital labour in the Global South (specifically, Brazil), arguing that informality constitutes its backbone. In economies where regulated markets never created enough jobs, informality is a structural feature of the division of labour. Digital labour reinforces longstanding historical inequalities by integrating such traditional practices into global technological outsourcing chains. Any policies must therefore address the profound asymmetries that perpetuate inequality, rather than vainly attempt to formalize informal labour.

What do social classes look like in (globalized) digital labour? Antonio A. Casilli addresses this question in the case of data workers. While terminological and conceptual heterogeneity may seem confusing at first sight, it is in fact a characteristic feature of class formation in progress.

It involves both precarious data workers and the tech middle class: what is emerging is a unified, meaningful political subject, even though its consolidation remains contested and fragmented.

### Misrecognised abilities in the knowledge economy

The third part engages with the social tensions that surround workers' abilities and their (lack of) recognition, disrupting their position in the labour market and potential future trajectories. By so doing, it also takes a stance with the idea that digital labour reconfigures the relationship of work with time – both daily time use, and the longer time of careers. Gemma Newlands introduces the topic by reviewing existing research on the three interconnected concepts of careers, employability, and perceived employability. She shows how, in digital labour, skills and experience do not usually match the way work histories are recognized and evaluated. This gap can undermine workers' perspectives in all three domains.

Marco Marrone advances this line of thought using the case of food delivery work in Italy. There is no such thing as algorithmic management replacing human know-how with automated processes: many informal and tacit skills play a role, from technical knowledge to time management and interpersonal abilities. But if algorithmic systems depend on the situated knowledge of workers, they obscure its value and contribution. This process is relational and rooted in organizational misrecognition rather than mere technological substitution.

Finally, Julieta Longo and Mariana Fernández Massi examine time use among creative freelancers in Argentina. They reveal how platforms individualize the way working time is defined, organized, and measured, leaving workers alone to reconstruct boundaries between work and non-work and to make decisions that were previously collective. They also uncover expanded margins of action, when workers interpret or repurpose technologies. Yet capacity to exert agency varies with experience and competencies, potentially deepening inequalities.

### Social inequalities

The last part of the book appraises the linkages between the emerging digital inequalities that arise from platforms and algorithmic management, and classical predictors of disparities, notably gender and race. Tammy Katsabian and Antonio Aloisi launch the reflection by examining anti-discrimination frameworks in the United States and the European Union, with focus on their strengths and weaknesses against digital technologies. They show that extant legal tools grapple to address proxy discrimination, data-driven correlations and the opacity of automated systems. Their proposed solution combines purposive interpretation of equality law, stronger use of data protection instruments, and enhanced collective participation.

Al James offers an introduction to a feminist research agenda. Platform architectures, algorithmic management and AI initially seemed to offer inclusive work environments – for example, by offering time and space flexibility. However, the 'analogue' constraints of motherhood and care remain. Inherited gendered relations continue to shape women's patterns of digital job search, working hours, remuneration, and career. These differences should now fully form part of the digital labour literature.

In the last chapter, Sophie Bernard analyses race in the case of Uber drivers in London, Montréal and Paris. Why, she asks, are most of them non-white? Her answer uses the platform business model, which needs a surplus workforce to face any demand peaks. Thus, the platform attracts drivers with promises of independence and high pay, and they see it as a better alternative to social and racial confinement to the 'subproletariat'. Under tight competition, though, workers find themselves relegated to it once again, as conditions and wages deteriorate.

## IMPLICATIONS AND CONCLUSIONS

Together, these analyses outline new directions for policymaking. First, stronger coordination between labour regulation and technology policy is necessary. The European Union and some other countries have laudably attempted to reintroduce some of the rights and protections associated to conventional salaried labour (though such achievements are always under threat from companies' lobbying); however, any policies that uncritically support innovation may unwittingly offset some of these efforts. For example, growing funding for AI production may lead firms to increase their demand of data work through platform intermediaries that partly escape labour legislation. Conversely, endeavours to curb technology (for example, through algorithmic transparency requirements) without considering its social and labour surroundings may prove ineffective.

Second, it is essential to think across borders and consider the impacts of changes in one jurisdiction throughout the planetary-scale interconnections that shape today's digital labour markets. For example, even enhanced workers' rights in the Global North may not suffice to narrow the overall gap between labour and capital if firms can offshore tasks to countries in which such protections are weaker. Real and more equitable improvements require coordinated strategies throughout the Global South to develop the local economy and retain the knowledge produced through digital labour.

Third, promotion of education and training does not suffice to guarantee fair working conditions and remunerations in the absence of adequate frameworks for the systematic recognition and valuation of workers' skills and experience. Even reskilling policies, explicitly aimed at endowing workers with digital competencies, do not protect against the effects of technologies that deteriorate working conditions. Prior learning recognition programmes, already developed in contexts of high informality, exemplify an alternative and more promising way forward.

Fourth, there is a need to prioritize the most vulnerable, by giving more consideration to the marginalized people and groups (including, but not necessarily limited to, women, non-whites, and migrants) who are less able to make the most of their digital labour experience and have fewer alternatives in the formal economy. This may be achieved through a mix of measures that, beyond labour regulation, include non-discrimination and data protection laws, appropriately reinterpreted to address the specificities of the digital age. Empowering workers' representatives – whether through established unions or grassroots initiatives – is also essential, though difficult when digital labour struggles cross national borders.

Overall, this book constitutes a step forward in the collective effort to reflect on the linkages and intersections between digital labour and the broad societal tendencies of today. Hopefully, the research avenues that it opens will deepen our social-scientific knowledge of current realities while also supporting the design and implementation of suitable policies.

## ACKNOWLEDGEMENTS

The author acknowledges financial support from ANR through grant ANR-23-CE38-0015.